# The Effect of an Under-Dense Plasma Density Gradient on the Backstreaming Ion Mechanism

George J. Caporaso
Lawrence Livermore National Laboratory, Livermore, California 94550 USA

*Abstract*

The space charge limited emission of ions from a target in the focus of an intense relativistic electron beam is studied analytically for the case of a spatially varying target density profile. In particular, the emission in the presence of an under-dense plasma shelf in contact with the solid density target dramatically differs from the case of an abrupt solid-vacuum boundary. It is found that an under-dense gradient scale length several times that of the beam radius at the focus reduces the emission by at least an order of magnitude over that to be expected from a solid-vacuum boundary.

## 1 INTRODUCTION

High-resolution x-ray radiography requires the production of a small (≈1 mm diameter) spot on the surface of a Bremsstrahlung converter target by a relativistic electron beam of at least several kiloamperes [1]. A mechanism that might possibly disrupt the focal spot was proposed by D. Welch [2]. Bombardment of the target by a high power electron beam would lead to the rapid formation of a surface plasma. A large axial electric field would appear at the surface due to the charge redistribution on the target arising from cancellation of the beam's *radial* electric field. This axial field would expel the ions into the beam. These *backstreaming* ions would acquire energies on the order of the space charge depressed potential of the beam and would propagate upstream at very high speeds where they would act as an electrostatic focusing lens. The focusing due to these moving ions would cause the electron beam to pinch upstream of the target and then rapidly diverge. The result would be a spot size that would rapidly increase in time at the converter target.

In the case of multiple pulses striking the same target the situation might be modified by the presence of a plasma left over from the previous pulses. The leading edge of this plasma could be very tenuous.

We consider the effects of the leading edge of the plasma, which is of lower density than the beam. In this region the plasma electrons will be expelled and a "bare" ion column will be present. We assume that the axial electric field is low enough that we can neglect the motion of these background ions. We treat emission of ions from the "critical surface" where the beam density is equal to the ion density and impose the space-charge limited flow condition.

An analytic model is presented for a "beer can" geometry in which a close fitting conducting tube surrounds the beam right up to the target.

The under-dense plasma is modeled as having an exponentially varying density.

A model of backstreaming ion emission from a sharp boundary was given previously [3]. That reference derived an analytic solution for the emission of ions from a planar target for the same geometry and discussed the subsequent disruption of the electron beam focal spot. This paper considers the more general case of a target that has a tenuous plasma in the vicinity of the target. The general solution is found as a function of the scale length of the under-dense plasma density gradient. In the limit that the scale length approaches zero the solution of reference [3] is recovered.[*]

Because the source of the electric field responsible for the axial flow of ions results from the rapid variation in radial electric field due to the neutralizing effects of the target, it is of interest to consider the effects of a neutralizing background that varies from zero density up to the density of the beam. By spreading out the region over which the radial electric field is neutralized it is expected that the axial electric field and hence the ion emission will be reduced.

## 2 TARGET GEOMETRY AND MODEL

Consider the geometry shown in figure 1. An electron beam that just fits inside a cylindrical conducting tube impinges normally on a conducting plane (target). We assume that the beam and tube have radius a and extend infinitely in the z-direction. We take the target to have a diffuse boundary with an exponentially varying density. We consider the case of steady-state space-charge-limited emission of ions.

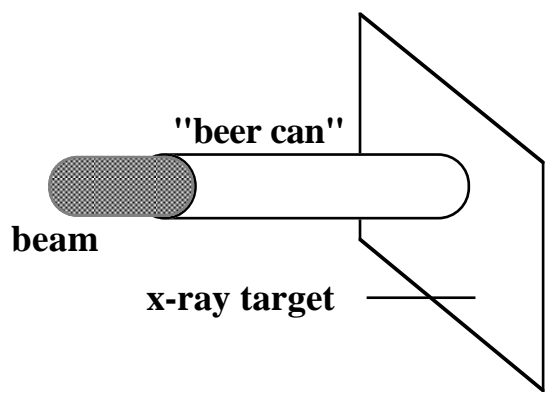

Fig. 1. "Beer can" geometry proposed to reduce the space charge depressed potential of the beam, which would reduce the backstreaming ion current.

---

[*] The value of the neutralization fraction given in equation [20] of reference [3] is in error.

We will assume that the target surface is sufficiently rich in ions that the flow will be space-charge-limited. The steady-state emission is determined by Poisson's equation for the electrostatic potential (in c.g.s. units)

$$\nabla^2 \Phi = -4\pi\rho = -4\pi\left[\rho_b\left(1 - f(z)\right) + \rho_i\right]. \quad [1]$$

Here f(z) represents the neutralization fraction of the beam's space charge ($\rho_b$) due to a static ion background. $\rho_i$ represents the charge density of the backstreaming ions. The ion velocity can be found from the conservation of energy (since the target and tube are grounded)

$$v_i = \sqrt{-2e\Phi / M} \quad [2]$$

where M is the ion mass and e is the ion charge. The (emitted) ion charge density is given by

$$\rho_i = J(r) / v_i \quad [3]$$

where J(r) is the ion current density.

Equation [1] is two dimensional (r and z). A great simplification is made possible by choosing the beam profile to be of the form

$$\rho_b = -\rho_o J_o(\alpha r) \quad [4]$$

where $J_O$ is the zeroth order Bessel function and $\alpha = x_{01} / a$. Here a is the radius of the beer can, $x_{01}$ is the first root of $J_O$ and $-\rho_O$ is the on-axis charge density of the beam.

Let us seek solutions which have the following form:

$$\Phi(r,z) = -\psi(z) J_o(\alpha r) \quad [5]$$

and

$$J(r) = \Lambda_o J_o^{3/2}(\alpha r) \quad [6]$$

where $\psi(z)$ and $\Lambda_o$ are to be determined.

Substitution of Equations [2] through [6] into Equation [1] yields

$$\psi - \frac{d^2\psi}{d\zeta^2} = -\frac{4\pi\rho_o[1 - f(\zeta)]}{\alpha^2} + \frac{4\pi\Lambda_o}{\alpha^2}\sqrt{\frac{M}{2e}}\frac{1}{\sqrt{\psi}} \quad [7]$$

where we have defined a dimensionless axial coordinate $\zeta=\alpha z$. We now define a dimensionless variable and a dimensionless constant

$$\Omega \equiv \frac{\alpha^2 \psi}{4\pi\rho_o} \quad [8]$$

and

$$\mu \equiv \frac{4\pi\Lambda_o}{\alpha^2}\sqrt{\frac{M}{2e}}\left(\frac{\alpha^2}{4\pi\rho_o}\right)^{3/2}. \quad [9]$$

We choose $f(\zeta)$ such that the critical surface occurs at z=0.

$$f(\zeta) = e^{-\zeta/\lambda}. \quad [10]$$

The parameter $\lambda$ is normalized gradient scale length. We would expect to recover previous results for a sharp boundary as $\lambda \to 0$. With these definitions the differential equation for the dimensionless potential becomes

$$\frac{\partial^2 \Omega}{\partial \zeta^2} - \Omega - \frac{\mu}{\sqrt{\Omega}} = e^{-\zeta/\lambda} - 1. \quad [11]$$

We use the boundary conditions appropriate for space-charge-limited flow arising from the critical surface (where the neutralization fraction is one):

$$\Omega(0) = 0 \qquad d\Omega(0)/d\zeta = 0. \quad [12]$$

As $\zeta \to \infty$ we have

$$\Omega_\infty + \frac{\mu}{\sqrt{\Omega_\infty}} = 1. \quad [13]$$

The asymptotic fractional neutralization of the beam due to the backstreaming ions is then just

$$f_n = \frac{\mu}{\sqrt{\Omega_\infty}}. \quad [14]$$

The differential equation [11] can be solved numerically and reveals that the potential changes very abruptly from zero at the target surface to the space charge depressed potential of the beam (as reduced by the backstreaming ions) over a distance of the order of the beam radius. When there is an under-dense plasma gradient the potential changes more slowly and rises to a higher value indicating less neutralization due to backstreaming ions.

The solution of Equation [11] is shown in Figure 2. Note that the potential changes rapidly over a distance of order the beam radius when no plasma is present.

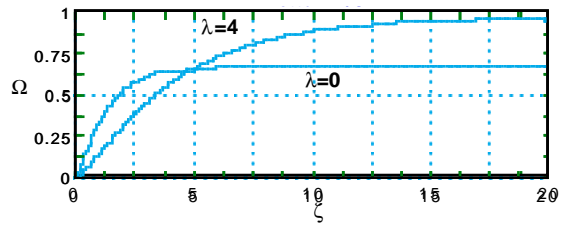

Fig. 2. Solution of Equation [11]. $\Omega$ is plotted vs. $\zeta$ for a sharp boundary ($\lambda$=0) and for one with a tenuous plasma shelf ($\lambda$=4).

The *normalized* neutralization fraction ($f(\lambda)/f(0)$) as a function of $\lambda$ is shown in Figure 3a. If the actual under-dense plasma scale length is L then the true neutralization fraction as a function of L/a is shown in Figure 3b.

## 3 Discussion

Figure 3b. shows that the neutralization fraction is decreased by an order of magnitude when the scale length is equal to a beam diameter. For typical parameters of interest for radiography the plasma blow-off speed is on the order of at least several mm/μsec. Thus, the scale length will be on the order of the beam diameter after several hundred nanoseconds. This time is sufficiently short that even though a plasma exists when a second beam pulse hits the target there may be no significant backstreaming ion emission.

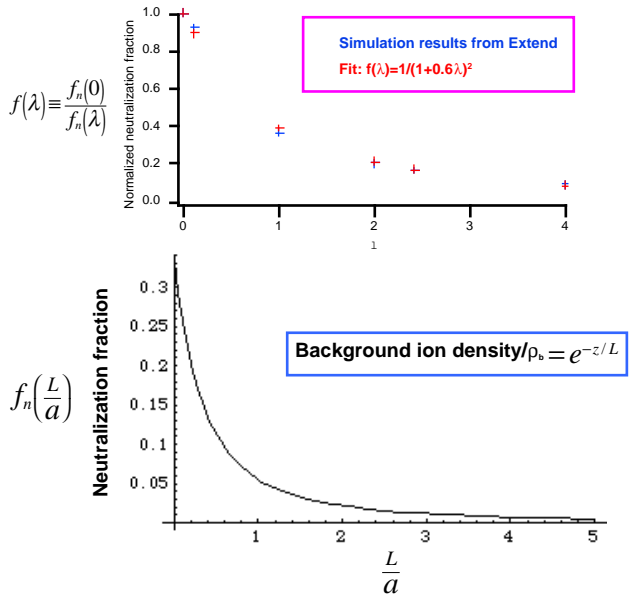

Fig. 3(a). Normalized neutralization fraction as a function of λ. 3 (b). The actual neutralization fraction as a function of L/a where L is the actual under-dense plasma scale length and a is the radius of the beam.

## 4 CONCLUSIONS

We have provided a solution to the problem of the space charge limited flow of ions from the critical surface of a target in the presence of an exponentially varying under-dense ion background. Modest density scale lengths are shown to substantially reduce the axial electric field and hence the emission of backstreaming ions from the target. This result indicates that the backstreaming ion mechanism may not be a serious threat to multiple pulse trains that hit a Bremsstrahlung target.

## 5 ACKNOWLEDGMENTS